\documentclass[
 ,final            
  ]
  {aipproc}

\layoutstyle{6x9}
\newcommand{\lesssim}{\lower.5ex\hbox{$\; \buildrel < \over\sim \;$}}
\newcommand{\gtrsim}{\lower.5ex\hbox{$\; \buildrel > \over\sim \;$}}
\newcommand{\e}{\epsilon}

\newcommand{\lC}{\bar\lambda_{\rm C}}
\def\E{{\cal E}}

\begin{document}

\title{On Hadronic Models for the Anomalous $\gamma$-ray 
Emission Component in GRB 941017}

\author{Charles D.\ Dermer}{
  address={Code 7653, Naval Research Laboratory, 
Washington, DC 20375-5352 USA}
}

\author{Armen Atoyan}{
  address={Centre de Recherche Math\'ematiques, 
Universite d\'e Montr\'eal, Montr\'eal, Canada H3C 3J7}
}

\begin{abstract}
Gonz\'alez {\it et~al}.\ (2003) have reported the discovery of an
anomalous radiation component from $\approx 1$ -- 200 MeV in GRB
941017.  This component varies independently of and contains $\gtrsim
3\times$ the energy found in the prompt $\sim 50$ keV -- 1 MeV
radiation component that is well described by the relativistic
synchrotron-shock model. Acceleration of hadrons to very high energies
by GRBs could give rise to a separate emission component. Two models,
both involving acceleration of ultra-high energy cosmic rays with
subsequent photomeson interactions, are considered. The first involves
a pair-photon cascade initiated by photohadronic processes in the GRB
blast wave. Calculations indicate that the cascade produces a spectrum
that is too soft to explain the observations. A second model is
proposed where photopion interactions in the GRB blast-wave shell give
rise to an escaping collimated neutron beam. The outflowing neutrons
undergo further photopion interactions to produce a beam of
hyper-relativistic electrons that can lose most of their energy during
a fraction of a gyroperiod in the Gauss-strength magnetic fields found
in the circumburst medium. This secondary electron beam produces
a hard synchrotron radiation spectrum that could explain the anomalous
component in GRB 941017.
\end{abstract}

\maketitle

  
\section{Introduction}

Based on joint analysis of BATSE LAD and EGRET TASC data, Gonz\'alez
et al.\ (2003) recently reported the detection of an anomalous MeV
emission component in the spectrum of GRB 941017 that decays more
slowly than the prompt emission detected with the BATSE LAD in the
$\approx 50$ -- 300 keV range. The multi-MeV component lasts for
$\gtrsim 200$ seconds (the $t_{90}$ duration of the lower-energy
component is 77 sec), and is detected with the BASTE LAD near 1 MeV
and with the EGRET TASC between $\approx 1$ and 200 MeV. The spectrum
is very hard, with a photon number flux $\phi(\e)\propto \e^{-1}$,
where $\e = h\nu/m_ec^2$ is the observed dimensionless photon energy.

This component is not predicted or easily explained within the
standard leptonic model for GRB blast waves, though it possibly could
be related to Comptonization of reverse-shock emission by the forward
shock electrons \cite{gg03}, including self-absorbed reverse-shock
optical synchrotron radiation \cite{pw03}. Another possibility is that
hadronic acceleration in GRB blast waves could produce this component.

We consider two models involving acceleration of hadrons at the
relativistic shocks of GRBs. In the first model, the anomalous
radiation component would be explained by the pair-photon cascade
radiation initiated by photohadronic processes between high-energy
hadrons accelerated in the GRB blast wave and the internal synchrotron
radiation field. We find, however, that the cascade radiation spectrum
is too soft to be able to explain the anomalous emission component of
GRB 941017.  In the second model, photomeson interactions in the
relativistic blast wave produce a beam of ultra-high energy neutrons,
as proposed for blazar jets \cite{ad03}. Photopion production of these
neutrons with photons outside the blast wave produce a directed
hyper-relativistic electron-positron beam in the process of charged
pion decay and the conversion of high-energy photons from $\pi^0$
decay. These energetic leptons produce a synchrotron spectrum in the
radiation reaction-limited regime extending to $\gtrsim$ GeV energies,
with properties in the 1 -- 200 MeV range similar to that measured
from GRB 941017. If this model is correct, detection of this component
therefore gives important indirect evidence for the acceleration of
ultrahigh energy cosmic rays in GRB blast waves.

\section{ Hadronic Models for GRB 941017}

We assume that efficient proton acceleration to ultra-high energies
takes place in GRB blast waves, as required in models where GRBs
accelerate high-energy cosmic rays \cite{vie95,wax95,der02}.
Photopion interactions of these protons with internal
synchrotron photons and with photons from an external radiation field
 create neutral particles and charged pions that 
decay and initiate an electromagnetic cascade within the GRB blastwave. 

\begin{figure}[t]
\vspace*{15.0mm} %
\includegraphics[width=7.0cm]{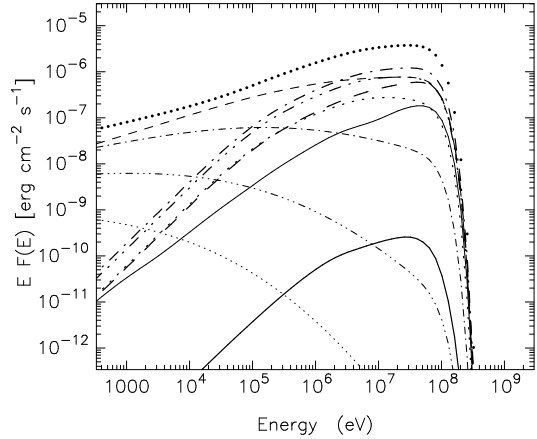}
\caption{Photon energy fluence from an electromagnetic cascade initiated
by photopion secondaries in a model GRB, with parameters given in the
text. Five generations of Compton (heavy curves) and synchrotron
(light curves) are shown. The first through fifth generations are
given by solid, dashed, dot-dashed, dot-triple--dashed, and dotted
curves, respectively. The total cascade radiation spectrum is given by
the upper bold dotted curve. }
\label{fig1}
\end{figure}

One possible model is to attribute the anomalous component to
radiation of the cascade produced within the blast wave.  Fig.\ 1
shows the hadron-initiated cascade radiation for a model GRB at
redshift $z =1 $, with hard X-ray fluence $\Phi_{tot} = 3\times
10^{-4}\,\rm erg \; cm^{-2}$, a light curve of 100 second duration
divided into 50 pulses of 1 second each, and Doppler factor $\delta =
100$ (see \cite{da03,ad03} for more details about the model). The
total amount of accelerated proton energy $E^\prime = 4\pi d_L^2
\Phi_{tot} \delta^{-3} (1+z)^{-1}$ is injected into the comoving frame
of the GRB blast wave.

The various generations of synchrotron and Compton radiation initiated
by the cascade are shown in Fig.\ 1, along with the total radiation
spectrum. As can be seen, the isotropic cascade radiation approaches
the spectrum of an electron distribution cooling by synchrotron
losses, that is, a spectral index between $-1.5$ and $-2$. This is too
soft to explain the observations of GRB 941017 with a spectral index
$= -1$. Moreover, the isotropic cascade radiation should decay at the
same rate as the synchrotron radiation. Thus the radiation from an
electromagnetic cascade in the GRB blast wave cannot explain the
anomalous component in GRB 941017.

We propose a second possibility based on our neutral beam model
\cite{ad03}. Ultra-relativistic neutrons formed in the reaction
$p+\gamma\rightarrow n +\pi^0$ are not confined by the magnetic field
in the GRB blastwave shell and flow out to create an energetic neutron
beam. These neutrons are subject to further photopion processes with
photons in the surrounding medium to form charged and neutral pions.
In the Gauss-type magnetic fields surrounding GRB sources that we
assume here, charged $\pi$ and $\mu$ at energies $\lesssim 10^{18}$ eV
decay rather than lose energy through synchrotron emission
\cite{rm98}. The charged pions decay into ultrarelativistic electrons
and neutrinos, whereas the decay of $\pi^0$ produces two $\gamma$ rays
that are promptly converted into electron-positron pairs on this same
external radiation field. These energetic electrons (including
positrons) are initially produced in the direction of the GRB jet.

The spectra of secondary electrons created by the neutron beam
displays a sharp cutoff at energies $\lesssim 10^{14}$ eV as a
consequence of the high threshold for photomeson interactions (see
Ref.\ \cite{da03} for more details). Electrons with Lorentz factor
$\gamma$ lose energy through synchrotron radiation in an ordered
magnetic field with strength $B$ at the rate $-d\gamma/dt =
\sigma_{\rm T} B^2 \gamma^2\sin^2\psi/(4\pi m_e c)$, where $\psi$
is the electron pitch angle. The corresponding synchrotron energy-loss
time scale $t_{syn} = \gamma/|d\gamma/dt|$. The gyration frequency
$\omega_B = eB/m_e c \gamma$, and is independent of pitch angle. When
$\omega_B t_{syn} \ll 1$, the electron loses almost all of its energy
into synchrotron radiation in a time less than the gyroperiod.  We use
the term ``hyper-relativistic'' to refer to electrons in this
radiation-reaction regime of synchrotron emission \cite{nw91,ab02}.
The mean energy of synchrotron photons from electrons that enter the
hyper-relativistic regime is independent of magnetic field (e.g.,
\cite{jag96}).

Electrons which cool before rotating by an angle equal
to the jet opening angle $\theta_j$ will produce
synchrotron photons that are primarily directed within the jet
opening angle. This condition is defined by $\omega_B t_{syn} \leq
\theta_j$, which applies to electrons with $\gamma
 \gtrsim \gamma_{hr}(\theta_j) = 
\sqrt{4\pi e/(\theta_j \sigma_{\rm T} B \sin^2\psi )}$ $\cong 3\times 10^7/
[\sin\psi\sqrt{(\theta_j/0.1) B({\rm G})}]$.  Here
we have taken a typical jet opening half-angle $\theta_j = 0.1$
because this is the average value implied by the analysis
leading to the standard energy reservoir result of GRBs \cite{fra01}.
Lower-energy electrons with  $\gamma < \gamma_{hr}(\theta_j)$
radiate their energy over a much larger solid angle and longer time.

The characteristic synchrotron photon energy $\E_\gamma = m_ec^2 \epsilon$
 radiated by electrons
which lose their energy within the jet opening angle $\theta_j$ is
\begin{equation}
 \E_j  \cong {\hbar e B\sin\psi\over m_e c}\; 
{ \gamma_{hr}^2(\theta_j) \over  (1+z) }\cong { 500 \over
(\theta_j/ 0.1) [(1+z)/ 2]\sin \psi }\;\;{\rm MeV}\;.
\label{Ej}
\end{equation}
Hyper-relativistic electrons with $\gamma > \gamma_{hr}(\theta_j)$
rapidly lose energy through synchrotron losses and deposit all of
their energy along the direction of the jet. Electrons at lower
energies are deflected to angles $\theta > \theta_j$, and their
emission is not seen by an on-axis observer. Hence the distribution of
electrons along the jet direction always has an effective low-energy
cutoff at $\gamma_{hr}(\theta_j)$. The production spectrum of the
electrons can have an intrinsic cutoff $\gamma_{co}$ due either to the
low-energy cutoff in the escaping neutron spectrum, or to the
neutron-induced photomeson secondary spectrum outside the GRB blast
wave. If $\rho
\equiv\gamma_{co}/ \gamma_{hr}(\theta_j)\geq 1$, then the observed
synchrotron spectrum is a power law with $-1.5$ index for $
\E_j  \lesssim \E_\gamma \lesssim \rho^2
 \E_j $, and a photon spectrum with the same spectral
index as the accelerated protons and escaping neutrons at photon
energies $\E_\gamma \gtrsim \rho^2  \E_j $ \cite{nw91}.
If $\rho < 1$, then the observed photon spectrum at $\E_\gamma \gtrsim
\E_j $ has the same spectral index as the
primary neutrons.
 
At photon energies $\E_\gamma \ll  \E_j $, the
observed spectrum is produced by the same hyper-relativistic electrons
with $\gamma \gtrsim \gamma_{hr}(\theta_j)$,
but at energies $\e $ well below the peak energy $ 3\gamma^2 \e_B$, where 
$\e_B \equiv B/B_{cr}$ and $B_{cr} = 4.41\times 10^{13}$ G 
is the critical magnetic field. We now derive this spectrum.

The differential energy radiated per dimensionless energy interval
$d\e$ per differential solid angle element $d\Omega$ in the direction
$\theta$ with respect to the direction of an electron moving with
Lorentz factor $\gamma$ is given by
\begin{equation}
{dE\over d\e d\Omega} = {e^2\over 3\pi^2\lC}
({\e \over \gamma \e_B})^2 (1+\gamma^2\theta^2)^2
(\Lambda_\parallel +
\Lambda_\perp)\;,
\label{dEdedO}
\end{equation}
where $\lC=\hbar/m_ec= 3.86\times 10^{-11}$ cm is the electron Compton
wavelength, and $\Lambda_\parallel = K_{2/3}^2(\xi)$ and
$\Lambda_\perp =
(\gamma\theta)^{2}K_{1/3}^2(\xi)/[1+(\gamma\theta)^{2}]$ are factors
for radiation polarized parallel and perpendicular to the projection
of the magnetic field direction on the plane of the sky defined by the
observer's direction \cite{jac75}.  The factor $\xi =
\e/\hat\e$, where $\hat \e =
3\e_B\gamma^2/(1+\gamma^2\theta^2)^{3/2}$, and $K_n(x)$ is a modified
Bessel function of the second kind, with asymptotes
$K_n(x) \rightarrow {1\over 2}\Gamma(n )(2/x)^n$ in the limit $x \ll
1$, and $K_n(x) \rightarrow\sqrt{\pi/2x}~\exp(-x)$ in the limit $x \gg
1$. The condition $\xi \ll 1$
corresponds to $\e \ll \hat\e$ where $K_n(\xi )$ are in their 
power-law asymptotes, and $\xi \gtrsim 1$
or $\e \gtrsim \hat \e$ is where $K_n(\xi )$ are in
exponential decline. The characteristic energy $\hat \e$ approaches
 $3\e_B\gamma^2$ when $\gamma\theta \ll 1$, and $\hat
\e$ declines with $\theta$ according to the relation $\hat \e \cong
3\e_B\gamma^2/(\gamma\theta)^3$ when $\gamma\theta \gg 1$.  When
$\e\ll\hat\e$, then $\Lambda_\parallel \gg \Lambda_\perp$ and $dE/d\e
d\Omega = (dE_{syn}/d\e d\Omega)\simeq 3^{1/3}(1.07
e/\pi)^2(\gamma\e/\e_B)^{2/3}/\lC \propto \e^{2/3}$.  For a fixed 
value of $\e$, this emissivity
exponentially cuts off  when $\e\gtrsim 3\e_B/\gamma\theta^3$, or when
$\theta \gtrsim \theta_{max}=(3\e_B/\gamma\e)^{1/3}$.

The synchrotron emission spectrum in the limit $\E_\gamma \ll 
\E_j$, integrated over solid angle, is simply given
\begin{equation}
{dE\over d\e}\simeq 2\pi \int_0^{(3\e_B/\gamma\e)^{1/3}}d\theta\theta\;
({dE_{syn}\over d\e d\Omega})\simeq {3e^2\over \pi \lC}\propto \e^0\;.
\label{dEde}
\end{equation}
This differs from the energy index
$+1/3$ for the electrons in the classical regime averaged over a
complete orbit because, in this case, $d\theta\sin\theta \rightarrow
d\theta \sin\psi $ in the integration in eq.\ (\ref{dEde})
\cite{rl79}.

\section{Neutron Beams from GRBs}

In our model, ultrarelativistic protons undergo photomeson
interactions with the internal and external radiation photons
\cite{da03}, producing a beam of outflowing neutrons.  
Subsequent interactions of these neutrons with the external radiation
field generates a beam of hyper-relativistic electrons. (The external
photons could be due to pre-supernova thermal or GRB synchrotron
radiations scattered by the stellar wind in the collapsar model, or by
plerionic emission in the SA model.) The synchrotron radiation has a
low-energy cutoff at $\E_\gamma < \E_j \sim 500$ MeV -- 1 GeV, with a
specific characteristic power-law number spectral index equal to $-1$,
as observed for GRB 941017 \cite{gon03}.

The $\approx 200$ s decay time of the anomalous emission can be
explained by the emission of hyper-relativistic electrons from the
edges of a jet blastwave at distances $R \approx
10^{15}(\theta_j/0.1)^{-2}[(1+z)/2]^{-1}$ cm, implying significant
opacity to photomeson processes due to external photons at these
distances.  The differing GRB external radiation and density
environments which determine the intensity of target photons could
account for the unusual spectrum of GRB 941017. The neutron beam can
carry up to $\sim 50$\% of the entire energy of injected protons
\cite{ad03}, which can be further reprocessed about equally into
neutrinos and hyper-relativistic electrons. This nevertheless requires
a large nonthermal hadron-to-lepton load in the GRB blast wave in
order for the fluence of the anomalous component to be comparable with
that of the normal component. This large hadron load is also
consistent with predictions linking cosmic rays to GRB sources
\cite{wda03}.

Our model predicts that most of the energy of hyper-relativistic
electrons will be observed at $\gtrsim 500$ MeV -- GeV energies as a
delayed radiation component, well-suited for observations with {\it
GLAST}.  GRBs with anomalous $\gamma$-ray emission components should
also be bright neutrino sources detectable with {\it IceCube}. 


\vskip 2.0mm
\noindent {\bf Acknowledgments.~} Discussions with M. Gonz\'alez, 
B.\ Dingus, A.\
K\"onigl and D.\ Lazzati are gratefully acknowledged.  This work is
supported by the Office of Naval Research and the NASA {\it GLAST}
program.

\end{document}